\documentclass[12pt]{article}
\makeatletter
\ifcase\@ptsize
 \font\teneufm=eufm10
 \font\seveneufm=eufm7
 \font\fiveeufm=eufm5
 \font\teneusm=eusm10
 \font\seveneusm=eusm7
 \font\fiveeusm=eusm5
\or
 \font\teneufm=eufm10 scaled \magstephalf
 \font\seveneufm=eufm7
 \font\fiveeufm=eufm5
 \font\teneusm=eusm10 scaled \magstephalf
 \font\seveneusm=eusm7
 \font\fiveeusm=eusm5
\or
 \font\teneufm=eufm10 scaled \magstep1
 \font\seveneufm=eufm7
 \font\fiveeufm=eufm5
 \font\teneusm=eusm10 scaled \magstep1
 \font\seveneusm=eusm7
 \font\fiveeusm=eusm5
\fi

\newfam\eufmfam
\newfam\eusmfam
\textfont\eufmfam=\teneufm  \scriptfont\eufmfam=\seveneufm
  \scriptscriptfont\eufmfam=\fiveeufm
\textfont\eusmfam=\teneusm  \scriptfont\eusmfam=\seveneusm
 \scriptscriptfont\eusmfam=\fiveeusm

\def\frak{\ifmmode\let\next\frak@\else
 \def\next{\errmessage{Use \string\frak\space only in math mode}}\fi\next}
\def\frak@#1{{\frak@@{#1}}}
\def\frak@@#1{\fam\eufmfam#1}

\def\sh{\ifmmode\let\next\sh@\else
 \def\next{\errmessage{Use \string\sh\space only in math mode}}\fi\next}
\def\sh@#1{{\sh@@{#1}}}
\def\sh@@#1{\fam\eusmfam#1}

\ifcase\@ptsize
 \font\tenmsa=msam10
 \font\sevenmsa=msam7
 \font\fivemsa=msam5
 \font\tenmsb=msbm10
 \font\sevenmsb=msbm7
 \font\fivemsb=msbm5
\or
 \font\tenmsa=msam10 scaled \magstephalf
 \font\sevenmsa=msam7
 \font\fivemsa=msam5
 \font\tenmsb=msbm10 scaled \magstephalf
 \font\sevenmsb=msbm7
 \font\fivemsb=msbm5
\or
 \font\tenmsa=msam10 scaled \magstep1
 \font\sevenmsa=msam7
 \font\fivemsa=msam5
 \font\tenmsb=msbm10 scaled \magstep1
 \font\sevenmsb=msbm7
 \font\fivemsb=msbm5
\fi

\newfam\msafam
\newfam\msbfam
\textfont\msafam=\tenmsa  \scriptfont\msafam=\sevenmsa
  \scriptscriptfont\msafam=\fivemsa
\textfont\msbfam=\tenmsb  \scriptfont\msbfam=\sevenmsb
  \scriptscriptfont\msbfam=\fivemsb

\def\Bbb{\ifmmode\let\next\Bbb@\else
 \def\next{\errmessage{Use \string\Bbb\space only in math mode}}\fi\next}
\def\Bbb@#1{{\Bbb@@{#1}}}
\def\Bbb@@#1{\fam\msbfam#1}
\def\hexnumber@#1{\ifnum#1<10 \number#1\else
 \ifnum#1=10 A\else\ifnum#1=11 B\else\ifnum#1=12 C\else
 \ifnum#1=13 D\else\ifnum#1=14 E\else\ifnum#1=15 F\fi\fi\fi\fi\fi\fi\fi}
\def\msa@{\hexnumber@\msafam}
\def\msb@{\hexnumber@\msbfam}
\mathchardef\square="0\msa@03

\makeatother
\newcommand{\beq}{\begin{equation}}
\newcommand{\eeq}{\end{equation}}
\newcommand{\ba}{\begin{array}}
\newcommand{\ea}{\end{array}}
\newcommand{\bea}{\begin{eqnarray}}
\newcommand{\eea}{\end{eqnarray}}
\newcommand{\bean}{\begin{eqnarray*}}
\newcommand{\eean}{\end{eqnarray*}}

\newtheorem{theorem}{Theorem}[section]
\newtheorem{prop}[theorem]{Proposition}

\newtheorem{remark}[theorem]{Remark}

\makeatletter
\@addtoreset{equation}{section}

\makeatother

\newcommand{\rref}[1]{(\ref{#1})} 

\def\endpf{\begin{flushright}$\square$\end{flushright}}

\begin{document}
\begin{titlepage}
\begin{center}
{\huge 
The Soliton Equations associated with the Affine Kac--Moody Lie Algebra 
 $G_2^{(1)}.$}
\end{center}
\vspace{0.8truecm}
\begin{center}
{\large
Paolo Casati$^\sharp$, Alberto Della Vedova$^\flat$,  Giovanni Ortenzi$^\odot$
\vskip  0.8truecm
$\sharp \odot$ Dipartimento di Matematica e Applicazioni\\
 Universit\`a di Milano-Bicocca\\
Via R. Cozzi, 53\\
20125 Milano, Italy\\
$\odot$ Dipartimento di Matematica\\
        Politecnico di Torino\\
        Corso Duca degli Abruzzi, 24 \\
        10129 Torino, Italy\\
$\flat$ Dipartimento di Matematica\\
                   Universit\`a di Parma\\
                   Viale G. P. Usberti, 53/A\\
                   43100 Parma, Italy }
\end{center}
E--mail: $\sharp$ casati@matapp.unimib.it \\
\phantom{E--mail: }$\flat$ alberto.dellavedova@unipr.it \\
\phantom{E--mail: }$\odot$ giovanni.ortenzi@unimib.it \\
\vspace{0.2truecm}
\vspace{0.2truecm}

\abstract{\noindent
We construct in an explict way the soliton equation corresponding to the 
affine Kac--Moody Lie algebra $G_2^{(1)}$ together with their bihamiltonian
structure. Moreover the Riccati equation satisfied by the generating 
function of the commuting Hamiltonians densities is also deduced. Finally we describe a  way to deduce
the bihamiltonian equations directly  in terms of this latter functions }\\
\end{titlepage}

\section{Introduction}
One of the most fascinating  discoveries of the last decades is surely 
the deep and fundamental link between the affine Kac--Moody
Lie algebras (and their groups as well),  and the soliton equations.
This relation  first studied and described under different points of view
in a sequel of seminal papers by Sato \cite{S1,SS} Date, Jimbo,  
Kashiwara and Miwa \cite{DJKM}, Hirota \cite{H} 
Drinfeld and Sokolov \cite{DS} and Kac and Wakimoto \cite{KW} has inspired  
almost innumerable further investigations and generalizations (see 
for example the quite interesting papers of Burroughs, de Groot, Hollowood, Miramontes
\cite{BdGHM} \cite{BdGHM1}).
Nevertheless, as far as we know,   it seems that  no explicit 
description of the hierarchy corresponding in the scheme of Drinfeld 
and Sokolov to the affine Kac--Moody Lie algebra $G^{(1)}_2$
(even of the first non trivial equations) can be found in the 
literature, fact probably related to the  size of standard realization of 
$G_2$ (namely by $7\times 7$ matrices). The aim of this letter 
is to fill this gap and to show how the bihamiltonian 
formulation  of the Drinfeld-Sokolov reduction \cite{CMP1} \cite{CP}
\cite{CFMP2} makes the computations involved more reasonable.
The main ingredient of our construction will be indeed the technique of the 
transversal submanifold, which can be implemented only in the bihamiltionian
reduction theory, and which drops drastically the free variables involved in the computations.
The same technique provides also a way to construct a Riccati type equation for the formal Laurent
series for the conserved quantities of the corresponding integrable system. Since this equation at least 
in principle may be iteratively solved in a pure algebraic way, the bihamiltonian technique offers a computational way
to construct the whole hierarchy. Moreover what happens in the case of the affine Lie algebras $A_n^{(1)}$ suggests 
that it could be exists  a way to obtain
directly the equations of the hierarchy,starting by such conserved quantities,
 without referring directly to the underlying bihamiltonian structure.\par
The paper is organized as follows in the first section we perform the bihamiltonian reduction of the Drinfeld-Sokolov 
hierarchy defined on the affine Kac--Moody Lie algebra $G_2^{(1)}$ obtaining the reduced bihamiltonian structures and the 
first equations of the hierarchy as well. In the second and last section we explain and perform the so called Frobenius 
technique for the same algebra obtaining a so called Riccati equation satisfied by the generating function of the 
conserved densities. Finally we shall show how the knowledge of this function is enough to construct the entire hierarchy
of bihamiltonian equation.\\\\
{\bf Aknowledgements} We would like to thank Marco Pedroni and Youjin Zhang for many useful discussions on the subject. 

\section{The Bihamiltonian Reduction Theory of the Lie algebra $G^{(1)}_2$ }
The aim of this first section is to obtain the bihamiltonian structure of the soliton equation associated with 
the Kac--Moody affine Lie algebra $G^{(1)}_2$ in the Drinfeld--Sokolov theory by performing a bihamiltonian reduction 
process.\par
For the convenience of the reader, let us start by recalling  the main facts of the bihamiltonian reduction theory,
referring for more details to the papers \cite{CMP1} \cite{CP} \cite{CFMP2} where this theory was first developed.
A bihamiltonian manifold $\cal M$ is a manifold equipped with two compatible Poisson structures, i.e., two Poisson 
tensors $P_0$ and $P_1$ such that the pencil $P_\lambda=P_1 - \lambda P_0$ is a Poisson tensor for any 
$\lambda \in{\Bbb C}$. Let us fix a symplectic submanifold $\cal S$ of $P_0$ and consider the distribution
$D=P_1 \mbox{Ker}(P_0)$ then  the bihamiltonian structure of $\cal M$, provided that the quotient
space $\cal N={\cal S}/E$ is a manifold, can be reduced on $\cal N$ (\cite{CMP1} Prop 1.1).\par    
To construct the reduced Poisson pencil $P^\mathcal{N}_\lambda$ from the Poisson pencil $P_\lambda$ on $\mathcal{M}$
we have to perform the following steps \cite{CFMPtr}:
\begin{enumerate}
\item For any 1--form $\alpha$ on $\mathcal{N}$ we consider the 1--form  $\alpha^*$ on $\mathcal{S}$, which obviously
belongs to the annihilator $E^0$ of $E$ in $T^*\mathcal{S}$.
\item We construct a 1--form $\beta$ on $\mathcal{M}$ which belongs to the annihilator $D^0$ of $D$ and satisfies
\beq
\label{lift}
i^*_\mathcal{S} \beta=\pi^*\alpha
\eeq
(i.e., a lifting of $\alpha$).
\item We compute the vector field $P_\lambda\beta$, which turns out (see \cite{CFMPtr} Lemma 2.2) to be tangent
to $\mathcal{S}$.
\item We project $P_\lambda\beta$ on  $N$:
$$
(P^\mathcal{N}_\lambda)_{\pi(s)}\alpha=\pi_*(P_\lambda)_s\beta.
$$
\end{enumerate}
We shall not compute in the next section  the reduced bihamiltonian structure related to the affine 
Kac--Moody Lie algebra
$G^{(1)}_2$ using directly the above cited Theorem but rather implementing the technique of the
transversal submanifold \cite{CP} and \cite{CFMPtr} in order to avoid most of the computations involved.\par
A transversal submanifold to the distribution $E$ is a submanifold $\mathcal{Q}$ of  $\mathcal{S}$, which intersects
every integral leaves of the distribution $E$ in one and only one point. This condition implies the following relations
on the tangent space:
\beq
\label{trans}
T_{q}\mathcal{S}=T_{q}\mathcal{S}\oplus E_q\qquad \forall q\in \mathcal{Q}
\eeq
The importance of the knowledge of a transversal submanifold lies in the following Theorem proved in\cite{CMP1}
\cite{CFMP2}: 
\begin{theorem}\label{trastec} 
Let $\mathcal{Q}$ be a transversal submanifold of  $\mathcal{S}$ with respect the distribution $E$.
Then $\mathcal{Q}$ is a bihamiltonian manifold isomorphic to the bihamiltonian manifold $\mathcal{N}$ and the 
corresponding reduced Poisson pair  is given by:
\beq
\label{Poispr}
\left(P_i^{\mathcal{Q}}\right)_q\alpha=\Pi_*(P_i)_q\tilde{\alpha})\qquad i=0,1
\eeq
where $q \in \mathcal{Q}$, $\alpha \in T^*_q \mathcal{Q}$  $\Pi_*:T_q\mathcal{S}\to T_q\mathcal{Q}$ is the projection
with respect the decomposition \rref{trans} and $\tilde{\alpha}\in T^*_q \mathcal{M}$ satisfies the conditions:
\beq
\label{formcond}
 \tilde{\alpha}_{\vert D_q}=0  \qquad \tilde{\alpha}_{\vert T_q\mathcal{Q}}=\alpha.
\eeq
\end{theorem}
Actually for our porpoises the hypothesis of this Theorem  may be slightly relaxed by considering 
a submanifold $\mathcal{Q}$ which is only locally transversal (i.e., it satisfies  only the  weaker 
condition \rref{trans}) in this case of course $\mathcal{Q}$ and $\mathcal{N}$ could  be  only locally isomorphic
(see \cite{LP} for more details).\par
The bihamiltionian manifolds which are interesting in, are  the bihamiltonian manifold 
naturally defined on the affine Kac--Moody Lie algebras.
An affine non twisted Lie algebra $\widehat{\frak g}$ can be realized 
 as a semidirect product of the central extensions of a loop algebra of 
a simple finite dimensional Lie algebra $\frak g$ and a derivation $d$:
$$
 \widehat{\frak g}=C^\infty(S^1,\frak g) \oplus{\Bbb C} d \oplus {\Bbb C} c.
$$
Then the Lie bracket of two (typical) elements 
in $\widehat{\frak g}$ of the type 
$$
X=x_n\otimes x^n+\mu_1c+\nu_1d, \qquad Y=x_m\otimes x^m+\mu_2c+\nu_2d
$$
with $x_n,y_n\in \frak g$ and $n,m\in \Bbb Z$,
 $\mu_1,\mu_2,\nu_1,\nu_2\in \Bbb C$ is
\beq
\label{Lbr}
\left[X,Y\right] =\left[x_n,y_m\right]\otimes x^{n+m}+
(x_n,y_m)c\delta_{n+m,0}-n\nu_2x_n\otimes x^n+
m\nu_1y_m\otimes x^m
\eeq
where $\left[x_n,y_m\right]$ is the Lie bracket in 
${\frak g}$, and $(\cdot,\cdot)$ is the killing form
of  ${\frak g}$. (and finally $\delta$ is the usual Kronecker delta).
In what follows the derivation $d$ will not play any role. Being $\widehat{\frak g}$ a affine (infinite dimensional) 
manifold we may identify it with its tangent space at any point.  Moreover using the non degenerated form
\beq
\label{notdeg}
\langle (V_1,a),(V_2,b) \rangle + \int_S^1 (V_1(x),V_2(x))dx+ab
\eeq
we may identify at any point $S$ the tangent space with the corresponding cotangent space 
$T_S \mathcal{M} =T^*_S \mathcal{M}$. Using these identifications we can write the canonical 
Lie Poisson tensor of $\widehat{\frak g}$ as
\beq
\label{canPt}
P_{(S,c)}(V)=c\partial_xV+\left[S,V\right].
\eeq
It can be easily shown that this Poisson tensor is compatible with constant Poisson tensor obtained by freezing
the tensor \rref{canPt} in any point of $\mathcal{M}$. In particular the hierarchies of Drinfeld and Sokolov turns out 
to be bihamiltonian with respect to the bihamiltonian pair $P_1$, $P_0$ where $P_1$ is the canonical Poisson tensor
\rref{canPt} and $P_0$ is the constant Poisson tensor  
\beq
\label{costP}
(P_0)_{(S,c)}(V)=\left[A,V\right].
\eeq
where $A$ is the constant function of $C^\infty(S^1,\frak g)$ whose value is the element of minimal weight in $\frak g$.

\section{The reduction process}
In this section, following \cite{CP}, we perform the bihamiltonian reduction of the exceptional Lie algebra $G_2^{(1)}$.
It is a rank 2 simple Lie algebra whose Cartan matrix is ${\left(\ba{cc} 2 & -3 \\ -1 & 2 \ea \right)}$. A possible Weyl
basis is:
\beq
\ba{cc}
H_1=d_{22}-d_{33}+d_{55}-d_{66} & H_2=d_{11}-d_{22}+2d_{33}-2d_{55}+d_{66}-d_{77} \\
E_1=d_{23}+d_{56} & E_2=d_{12}+d_{34}+2d_{45}+d_{67}\\
F_1=d_{32}+d_{65} & F_2=d_{21}+2d_{43}+d_{54}+d_{76}\\
\ea
\eeq
where $d_{ij}$ is a $7 \times 7$ matrix with $1$ in the $ij$ position and zero otherwise.\\
Thus the elements of  the algebra has the form
$$v=
\left[ \begin {array}{ccccccc} h_{{2}}&e_{{2}}&-e_{{3}}&2\,e_{{4}}&-6
\,e_{{5}}&6\,e_{{6}}&0\\\noalign{\medskip}f_{{2}}&h_{{1}}-h_{{2}}&e_{{
1}}&e_{{3}}&-2\,e_{{4}}&0&6\,e_{{6}}\\\noalign{\medskip}f_{{3}}&f_{{1}
}&-h_{{1}}+2\,h_{{2}}&e_{{2}}&0&-2\,e_{{4}}&6\,e_{{5}}
\\\noalign{\medskip}4\,f_{{4}}&-2\,f_{{3}}&2\,f_{{2}}&0&2\,e_{{2}}&-2
\,e_{{3}}&4\,e_{{4}}\\\noalign{\medskip}6\,f_{{5}}&-2\,f_{{4}}&0&f_{{2
}}&h_{{1}}-2\,h_{{2}}&e_{{1}}&e_{{3}}\\\noalign{\medskip}6\,f_{{6}}&0&
-2\,f_{{4}}&f_{{3}}&f_{{1}}&-h_{{1}}+h_{{2}}&e_{{2}}
\\\noalign{\medskip}0&6\,f_{{6}}&-6\,f_{{5}}&2\,f_{{4}}&-f_{{3}}&f_{{2
}}&-h_{{2}}\end {array} \right].
$$
As already noted on $G^{(1)}_2$ is  defined a bihamiltonian structure given by canonical Lie Poisson tensor 
and by the tensor \rref{costP} where in the present case the element of minimal weight is 
 $A=F_6$.\\
To perform the Marsden-Ratiu reduction process of such bihamiltonian structure we need to compute  
 first  a symplectic leaf $\mathcal{S}$ of $P_0$  and
second the  distribution $E=P_1(\mbox{Ker}(P_0)$ on the point of  $\mathcal{S}$.
 As proved in \cite{CP}  the symplectic leaves  of the 
constant Poisson tensor are affine subspaces modelled on the subspace of $G^{(1)}_2$ orthogonal to the isotropic
algebra of the element $A$. Following Drinfeld and Sokolov let us choose that passing
through the point $b=E_1+E_2$:
$$S=b+h(t)(2H_1+H_2)+f_1(t)F_1+f_3(t)F_3+f_4(t)F_4+f_5(t)F_5+f_6(t)F_6.$$
Then the  (constant) distribution $E=P_1 \ker P_0$ evaluated on the points of $\mathcal{S}$ is
$$
\left[ \begin {array}{ccccccc} t_{{1}}&0&0&0&0&0&0
\\\noalign{\medskip}0&t_{{1}}&0&0&0&0&0\\\noalign{\medskip}t_{{2}}&t_{
{3}}&0&0&0&0&0\\\noalign{\medskip}2\,t_{{4}}&-2\,t_{{2}}&0&0&0&0&0
\\\noalign{\medskip}t_{{5}}&-t_{{4}}&0&0&0&0&0\\\noalign{\medskip}t_{{
6}}&0&-t_{{4}}&t_{{2}}&t_{{3}}&-t_{{1}}&0\\\noalign{\medskip}0&t_{{6}}
&-t_{{5}}&t_{{4}}&-t_{{2}}&0&-t_{{1}}\end {array} \right]. 
$$
Luckily enough we may apply  Theorem \ref{trastec} since the submanifold $\mathcal{Q}$ of $\mathcal{S}$ 
$$
\mathcal{Q}=
\left[ \begin {array}{ccccccccccc} 0&1&0&&0&&0&&0&&0\\\noalign{\medskip}0&0&1
&&0&&0&&0&&0\\\noalign{\medskip}0&u_{{0}}&0&&1&&0&&0&&0\\\noalign{\medskip}0&0
&0&&0&&2&&0&&0\\\noalign{\medskip}0&0&0&&0&&0&&1&&0\\\noalign{\medskip}6\,u_{{
1}}&0&0&&0&&u_{{0}}&&0&&1\\\noalign{\medskip}0&6\,u_{{1}}&0&&0&&0&&0&&0
\end {array} \right]
$$
 is transversal to $E$.\\
The actually computations of  the explicit form of the reduced Poisson pencil as observed in \cite{CFMP3} 
 boils down  to find  (given a  1-form $v=(v_0,v_1)\in T^*(\mathcal{Q})$)  a
 section $V(v)$  in $i_{\mathcal{Q}}^*(T^*\mathcal{M})$ (where  $i_{\mathcal{Q}}\mathcal{Q} \hookrightarrow \mathcal M$
is the canonical inclusion) such that $P_{\lambda}\, V(v)\in T Q$. This implies that the entries of 
$V(v)$ are polynomials functions of the elements $(8\, e_1, 288\,e_6)$ and that the reduced Poisson pencil 
 $$
\frac{d\, q}{dt_{\lambda}} = (P_{\lambda}^Q)_q v = 
V(v)_x + [ V(v) + \lambda\, a, q],$$
becomes

\begin{eqnarray*} 
\frac{d\, u_0}{dt_{\lambda}} & = & 
\beta \left( 
\frac{7}{32} u_0' v_1^{(4)} 
+ \frac{3}{4} u_1 v_1' 
- \frac{1}{8} \left( u_0' \right) ^{2} v_1'
+ \frac{11}{32} u_0'' v_1''' 
+ \frac{5}{32} u_0^{(4)} v_1'
+ \frac {5}{16} u_0''' v_1'' 
+ \frac{1}{16} u_0 v_1^{(5)} + \right. \\ & & \left. 
+ \frac{5}{8} u_1' v_1 
- \frac{1}{32} u_0^2 v_1''' 
+ \frac{1}{32} u_0^{(5)} v_1 
+ \frac{1}{4} u_0 v_0'
+ \frac{1}{8} u_0' v_0 
- \frac{7}{4}  v_0''' 
- \frac{1}{32} v_1^{(7)} 
- \frac{1}{8} u_0 u_0'' v_1' + \right. \\ & & \left.
- \frac {5}{32} u_0 u_0' v_1'' 
- \frac {3}{32} u_0' u_0'' v_1
- \frac{1}{32} u_0 u_0''' v_1 \right)
-\lambda \frac{3}{4} v_1'
\end{eqnarray*} 

\begin{eqnarray*} 
\frac{d\, u_1}{dt_{\lambda}} & = & 
 \left(
\frac{1}{96} u_0' v_1^{(8)} 
- \frac{7}{144} (u_0''')^2 v_1'
- \frac{1}{1728} u_0^4 v_1'''
+ \frac{1}{1728} u_0^{(9)} v_1
- \frac{1}{32} u_0^2 v_0'''
+ \frac{3}{4} u_1 v_0' + \right. \\ & & \left.
+ \frac{1}{72} u_1 v_1^{(5)}
+ \frac{47}{576} u_0^{(4)} v_1^{(5)}
+ \frac{13}{144} u_1'' v_1'''
+ \frac{5}{288} u_0^2 u_0' v_1^{(4)}
+ \frac{7}{864} u_0 u_0' u_0^{(4)} v_1 + \right. \\ & & \left.
- \frac{61}{576} (u_0'')^2 v_1'''
+ \frac{1}{432} u_0 v_1^{(9)}
+ \frac{5}{96} u_1^{(4)} v_1'
+ \frac{1}{8} u_1' v_0
- \frac{1}{1728} v_1^{(11)}
- \frac{23}{576} u_0 u_0^{(5)} v_1'' + \right. \\ & & \left.
+ \frac{1}{16} u_0 v_0^{(5)}
+ \frac{29}{288} u_1''' v_1''
+ \frac{85}{1728} u_0^{(6)} v_1'''
+ \frac{1}{48} u_0^{(7)} v_1''
+ \frac{3}{32} u_0' v_0^{(4)}
- \frac{1}{24} u_1 u_0'' v_1' + \right. \\ & & \left.
- \frac{1}{24} u_0 u_1' v_1''
- \frac{1}{36} u_1 u_0 v_1'''
- \frac{1}{24} u_1 u_0' v_1''
+ \frac{23}{864} u_0'' u_0^2 v_1'''
+ \frac{29}{864} u_0 (u_0')^2 v_1''' + \right. \\ & & \left.
- \frac{17}{96} u_0'' u_0''' v_1''
- \frac{5}{64} u_0'' u_0^{(4)} v_1'
- \frac{1}{72} u_0'' u_0^{(5)} v_1
- \frac{7}{1728} u_0^2 (u_0')^2 v_1'
- \frac{67}{432} u_0' u_0''' v_1''' + \right. \\ & & \left.
- \frac{1}{48} u_0^{(4)} u_0''' v_1
- \frac{1}{32} u_0 u_0' v_0''
+ \frac{1}{144} u_0^2 u_1' v_1
+ \frac{1}{72} u_0^2 u_1 v_1'
- \frac{1}{432} u_0^3 u_0'' v_1'
- \frac{1}{1728} u_0^3 u_0''' v_1 + \right. \\ & & \left.
- \frac{1}{288} u_0^3 u_0' v_1''
- \frac{1}{1728} u_0 (u_0')^3 v_1
- \frac{11}{864} u_0 u_0{(6)} v_1'
- \frac{1}{576} u_0 u_0^{(7)} v_1
- \frac{5}{48} u_0' u_0^{(4)} v_1'' + \right. \\ & & \left.
- \frac{11}{288} u_0' u_0^{(5)} v_1'
- \frac{5}{864} u_0' u_0^{(6)} v_1 
- \frac{61}{864} u_0 u_0^{(4)} v_1'''
+ \frac{3}{32} u_0'' v_0'''
+ \frac{35}{576} u_0''' v_1^{(6)}
+ \frac{1}{64} (u_0')^3 v_1'' + \right. \\ & & \left.
- \frac{1}{32} v_0^{(7)}
+ \frac{35}{864} u_0 u_0' u_0''' v_1'
- \frac{1}{18} u_0 u_0'' v_1^{(5)}
- \frac{1}{36} u_0' u_1' v_1'
+ \frac{1}{72} u_0 u_0' u_1 v_1 + \right. \\ & & \left.
+ \frac{1}{32} u_0''' v_0''
+ \frac{5}{144} u_1' v_1^{(4)} 
- \frac{1}{288} u_0^2 v_1^{(7)}
+ \frac{1}{72} u_0 u_0'' u_0''' v_1
- \frac{1}{432} u_0^2 u_0' u_0'' v_1 + \right. \\ & & \left.
+ \frac{11}{144} u_0^{(5)} v_1^{(4)}
+ \frac{11}{144} u_0 u_0' u_0'' v_1''
- \frac{5}{64} u_0 u_0''' v_1^{(4)}
- \frac{5}{288} u_0 u_1''' v_1
- \frac{5}{288} u_0' u_1'' v_1 + \right. \\ & & \left.
- \frac{7}{144} u_0 u_1'' v_1'
- \frac{1}{72} u_1' u_0'' v_1
+ \frac{19}{576} (u_0')^2 u_0'' v_1'
+ \frac{13}{576} u_0^2 u_0''' v_1''
+ \frac{17}{1728} u_0^2 u_0^{(4)} v_1' + \right. \\ & & \left.
+ \frac{1}{96} u_0' (u_0'')^2 v_1
+ \frac{1}{36} u_0 (u_0'')^2 v_1'
+ \frac{13}{1728} u_0''' (u_0')^2 v_1
+ \frac{1}{576} u_0^2 u_0^{(5)} v_1
- \frac{1}{72} u_1 u_0''' v_1 + \right. \\ & & \left.
- \frac{7}{288} u_0 u_0' v_1^{(6)}
- \frac{5}{36} u_0' u_0'' v_1^{(4)} 
+ \frac{1}{96} u_1^{(5)} v_1 
+ \frac{1}{432} u_0^3 v_1^{(5)}
+ \frac{1}{192} u_0^{(8)} v_1' + \right. \\ & & \left.
+ \frac{1}{32} u_0'' v_1^{(7)} 
- \frac{7}{192} (u_0')^2 v_1^{(5)} 
\right)
- \lambda \left( 
- \frac{1}{72} u_0''' v_1
+ \frac{1}{72} u_0^2 v_1'
+ \frac{1}{72} v_1^{(5)}
- \frac{1}{24} u_0'' v_1' + \right. \\ & & \left.
+ \frac{1}{72} u_0 u_0' v_1 
- \frac{1}{24} u_0' v_1''
- \frac{1}{36} u_0 v_1'''
+ \frac{3}{4} v_0'
\right)
\end{eqnarray*} 
where the prime indicates the space derivative.
Having the reduced bihamiltonian structure 
we are now able  to write explicitly the first non-trivial flows of the hierarchy \cite{CMP1}. Since the Casimirs of 
$P_0$ are given by  
the functionals 
\beq
\label{kerP0}
H_0=\int_{S^1}dx \, u_0  \quad \mbox{and} \quad  H_1=\int_{S^1}dx \, u_0''u_0-\frac{u_0^3}{3}+108 u_1  
\eeq
we obtain
\bea
u_{0,t_0}&=&\frac{1}{8} u_{0x}\\
u_{1,t_0}&=&\frac{1}{8} u_{1x}
\eea
and
\bea
u_{0,t_1}&=&-\frac{1}{864}(u_0^{(5)}+5u_0'u_0^2-5u_0'''u_0-5u_0''u_0'-540u_1)\\
u_{1,t_1}&=&-\frac{1}{864}(-9u_1^{(5)}+15u_1'''u_0+15u_1''u_0'+10u_1'u_0''-5u_1'u_0^2)
\eea

\section{The Frobenius Technique}
In the first section we have found the bihamiltonian structure of the soliton equation associated to the affine Kac--Moody
Lie algebra $G_2^{(1)}$ together with its  first not trivial equations.
This is of course by far not the same thing as to provide a way to actually compute all the 
soliton equations of the hierarchy. In the setting of the bihamiltonian theory this second important
problem is tackled by looking for Casimirs of the Poisson pencil $P_\lambda=P_1-\lambda P_0$ i.e., solutions of the 
equations
\beq
\label{Caseq}
V_x+\left[V,S+\lambda A\right]=0\qquad s\in \mathcal{S}
\eeq 
which are formal Laurent series $V(\lambda)=\sum_{k=-1}^\infty V_k\lambda^{-k}$ 
whose coefficients are  one forms defined at least on the points of $\mathcal{S}$  
which are exact when restricted on $\mathcal{S}$. Indeed once such a solution is found the vector fields 
of the hierarchy can be written in the bihamiltonian form $X_k=P_0V_k=P_1V_{k-1}$.\par
This latter task is unfortunately usually a very tough problem, but in the contest of the integrable systems defined 
on affine Lie algebra   it can be solved by using the generalization of the dressing method of Zakharov Shabat 
proposed by Drinfeld and Sokolov \cite{DS}. 
Unfortunately exactly as happens for the Drinfeld--Sokolov reduction for the Lie algebra $G_2^{(1)}$ the 
computations involved to derive the explicit expression of the bihamiltonian fields of the hierarchy are very 
complicated. The aim  of this last section is to show how the so called
Frobenius technique (\cite{CFMPfract}) provides somehow a shortcut of the Drinfeld--Sokolov procedure.\par
More precisely this technique will give  a way to compute algebraically by a recursive procedure the 
conserved densities of the hierarchy and therefore the corresponding (maybe without passing through the 
Poisson tensors) bihamiltonian vector fields as well. However implementing such technique
requires  to give up the pure geometrical description of the hierarchy of the first section in order to   
consider also the minimal true loop module $C^\infty(S^1,{\Bbb R}^7)$ of $G^{(1)}_2$ \cite{CHPR} together with its 
geometrical dual space  and the set of its  linear automorphisms as well.\par
The starting point of the theory  is indeed to observe (\cite{CFMP3},) that $V\in T^*_S\mathcal{S}$ 
solves \rref{Caseq} at the point $S\in (\mathcal{S},c=1)$
if and only if it commutes viewed as linear operator 
in $\mbox{End}(C^\infty(S^1,{\Bbb C}^7))$ ( up the canonical identification
explained in the previous section) with the linear differential operator $-c\partial_x+S+\lambda A$.
Although this latter task seems  not really easier then the first one, 
it suggests a way (using the action of the affine Lie group $\widetilde{\frak G}$ on the representation space 
$C^\infty(S^1,{\Bbb R}^7))$
to obtain directly the equations of the hierarchy together with their hamiltonians.
Following what suggested by Drinfeld and Sokolov we can find the elements $V$ commuting with $-c\partial_x+S+\lambda A$
using the  observation that the element 
$B+\lambda A$  is a regular element and therefore its  isotropic subalgebra  ${\frak g}_{B+{\lambda} A}$  
is  a Heisenberg subalgebra $\frak H$ of $\widetilde{\frak g}$ 
spanned (up to the central charge) in our representation by the matrices 
$\Lambda^{6n+1}=\left(\frac{\lambda}{24}\right)^n(B+\frac{\lambda}{24} A)$,
 $\Lambda^{6n-1}=\left(\frac{\lambda}{24}\right)^{n-1}(B+\frac{\lambda}{24} A)^5$ with $ n\in \Bbb Z$. 
(For sake of simplicity, from now on, 
we rescale $\frac{\lambda}{24} \to \lambda$).
From this fact Drinfeld and Sokolov proves indeed  the
\begin{prop}\label{DSZS} For any operator of the form  $-\partial_x+S+\lambda A$ with $s\in \mathcal{S}$ there exists
a element $T$ in $\widetilde{\frak{G}}$ such that:
\beq
\label{DSSZeq}
T(-\partial_x+S+\lambda A)T^{-1}=\partial_x+(B+\lambda A)+H, \quad H\in \frak H.
\eeq
Therefore the set of the elements in $\widetilde{\frak g}$ commuting with $-\partial_x+S+\lambda A$ is given by
$T^{-1}{\frak H}T$. 
\end{prop}
The knowledge of a such a $T$  allows us to compute explicitly for any choice of an element in $\frak H$ the 
corresponding hierarchy of vectors fields together with their Hamiltonians.
\begin{prop}\label{gencassol} Let $C=\sum_{j=\pm 1\mathrm{mod}(6)} c_j\Lambda^{-j}$ with $c_j\in \Bbb C$
be an element in $\frak H$ then: 
\begin{enumerate}
\item the element $V_C=T^{-1}CT$ solves equation \rref{Caseq};
\item its hamiltonian on $\mathcal{S}$ is the function $H_C=\langle J,C\rangle$ where $J$ is defined by 
the relation 
\beq
\label{mommap}
J=T(S+\lambda A)T^{-1}+T_xT^{-1}
\eeq
\item in particular if $C$ has the form $C=\Lambda^j$, $j=\pm 1\mbox{mod}(6)$ (say $j=6n\pm 1$) 
then $V_C$ and $H_C$ simply denoted respectively $V^j$ 
has the Laurent expansion 
\beq
\label{LeVcHc}
V^j= \lambda^n\sum_{p\geq -2}\frac{1}{\lambda^{p+1}}V_{6p\pm 1} 
\eeq
\label{biheq}
\end{enumerate}
\end{prop}
{\bf Proof .} 
\begin{enumerate}
\item It was already proved above.
\item Using equation \rref{mommap} we can rewrite equation \rref{DSSZeq} in the form 
$T(-\partial_x+S+\lambda A)T^{-1}=-\partial_x+J$ showing the $J$ commutes with $C$ then:
$$
\frac{d}{dt}H_C=\langle \dot{J},C\rangle=\langle T\dot{S}T^{-1}+\left[\dot{T}T^{-1},J \right],C\rangle
$$
but since $C$ commutes with $J$ we get
$$
\frac{d}{dt}H_C=\langle T\dot{S}T^{-1},C\rangle=\langle \dot{S},T^{-1}C T\rangle=\langle \dot{S},V_C\rangle.
$$
\item Equation \rref{LeVcHc} follows for $j=6n+1$ from the identity
$$
\ba{c}
\mbox{res}(\lambda^{p-n}V^j)=\mbox{res}(\lambda^{p-n}T\Lambda^j T^{-1})
=\mbox{res}(\lambda^{p-n}T\lambda^n\Lambda T^{-1})  \\
=\mbox{res}(T\lambda^{p}\Lambda T^{-1})=\mbox{res}(T \Lambda^{6p+1}T^{-1})=\mbox{res}(V^{6p+1})
\ea
$$
while similar computations show that if $j=6n-1$ then $\mbox{res}(\lambda^{p-n}V^j)=\mbox{res}(V^{6p-1})$.\par
\end{enumerate}
\endpf

As already pointed out the actually computation of the element $T$ (which by the way provides also the vector fields
of the hierarchy) is in our case quite complicated. To avoid such computations let us first solve the related 
problem of finding  the  eigenvalues of the operator  $-\partial_x+S+\lambda A$:
\beq
\label{alp}   
-\psi_x+(S+\lambda A)\psi=\mu\psi.
\eeq
This latter problem can be  solved by the   observation that   the integral leaves $E$ are orbits of a group
action, completely characterized  by the distribution $E$ at the special point $B$. It holds indeed:
\begin{prop}\label{prop1F} The subspace ${\frak g}_{AB}:=\{V\in {\frak g}_{A}\vert V_x+\left[V,B\right]
\in{\frak g}_{A}^\perp\}$
is a subalgebra of $\frak g$ contained in the nilpotent  subalgebra ${\frak n}_-$ of loops with values in the maximal 
nilpotent subalgebra spanned by the negative (it depends how $G_2$ is defined). Therefore the corresponding group
$G_{AB}=\exp({\frak g}_{AB})$ is well defined. The distribution $E$ is spanned by the vector fields 
$(P_1)_B(V)$ with $V$ belonging to ${\frak g}_{AB}$, and its integral leaves are the orbits of the 
gauge action of $G_{AB}$ on $\mathcal{S}$ defined by:
\beq
\label{gauact}
S'=TST^{-1}+T_xT^{-1}.
\eeq  
\end{prop}
Explicitly the for the Lie algebra $G_2$ the group $G_{AB}$ is:
Now  it easily to see that equation \rref{gauact} implies on the space  $\mbox{End}(C^\infty(S^1,{\Bbb C}^7)$ that 
the linear differential operators $-\partial_x+S+\lambda A$  and $-\partial_x+S'+\lambda A$ are conjugated by an element
$T\in G_{AB}$in formula:
\beq
\label{guar7}
 (-\partial_x+S'+\lambda A)\circ T= T\circ(-\partial_x+S+\lambda A).
\eeq
 if $S$ and $S'$ satisfy \rref{gauact} with the same $T$.\par
Therefore if we define $v^{(0)}=(1,0,0,0,0,0,0)$ and by recurrence
\beq
\label{recrel}
v^{(j+1)}(S)=\partial_xv^{(j)}(S)+(S+\lambda A)v^{(j)}(S)\qquad (v^{(0)}(S)=v^{(0)})
\eeq
then we have the
\begin{prop}\label{recprop} The vectors $v^{(j)}(S)$ $j\geq 0$ are covariant 
i.e.,  $v^{(j)}(S')=v^{(j)}(S)T$ whenever  $S'=TST^{-1}-T_xT^{-1}$ with $T\in G_{AB}$. 
Moreover the subset $\{v^{(j)}\}_{j=0,\dots,6}$
is for any $S$ in $\mathcal{S}$ a basis for ${\Bbb C}^7$.
\end{prop} 
Developing now  the first dependent vector namely $v^{(7)}(S)$ we obtain the relation
\beq
\label{chareq}
\ba{rl}
v^{(7)}(S)=&2u_0v^{(5)}(S)+5u_{0}'v^{(4)}(S)+(6u_0''-u_0^2)v^{(3)}(S)+(4u_0'''-3u_0u_0')v^{(2)}(S)\\
&+(24u_1-4\lambda-(u_0')^2-u_0u_0''+u_0^{(4)})v^{(1)}(S)+12 u_1' v^{(0)}(S)
\ea
\eeq
called the characteristic equation, moreover   it is not difficult to show that, by construction,
 $u_0$ and $u_1$ are a complete set of invariants for the action of ${\frak g}_{AB}$ i.e., 
they can be used to parameterize the quotient space 
$\mathcal{N}$. \par
These covariant vectors are the main tool to solve the eigenvector problem stated above. It holds namely
\begin{prop}
\label{eigpro} If $\psi$ is the element of $C^\infty(S^1,{\Bbb C}^7)$ defined by the relations  
$\langle v^{(0)},\psi\rangle=1$, $\langle v^{(1)},\psi\rangle=h$ and  $\langle v^{(k)},\psi\rangle=h^{(k)}$
$k=2,\dots,6$ where the function $h^{(k)}$ are defined by the recurrence: $h^{(1)}=h$, $h^{(k+1)}=h^{(k)}_x+h^{(k)}h$
and $h$ satisfies  the ``Riccati''-type equation
\beq
\label{Riceqn}
\ba{rl}
h^{(7)}=&2u_0 h^{(5)}+5u_{0}'h^{(4)}+(6u_0''-u_0^2)h^{(3)}+(4u_0'''-3u_0u_0')h^{(2)}\\
&+(24u_1-4\lambda-(u_0')^2-u_0u_0''+u_0^{(4)})h+12 u_1' 
\ea
\eeq
then $\psi$ is an eigenvector of $-\partial_x+S+\lambda A$ with eigenvalue $h(z)$. 
Moreover equation \rref{Riceqn} admits a solution of the form $h(z)=cz+\sum_{i< 0}h_iz^{-i}$ where $z^6=\lambda$,
and  the coefficients $h_k$ are obtained iteratively in an algebraic way.
\end{prop}   
\begin{remark} Up to the transformation $h=\frac{\psi_x}{\psi}$ and the change of coordinates
$$
u_0=-u \qquad u_1=\frac{1}{12}(u^{(4)}-u''u-(u')^2-v)
$$
the equation \rref{Riceqn} coincides with the spectral problem arising from the Lax operator for $G_2^{(1)}$ 
given in \cite{DS}.
\end{remark} 
The Laurent expansion can be effectivley computed using \rref{Riceqn} its first term are:
\beq
\label{solRic}
\ba{rl}
c &= 2^{1/3} \\
h_{0}&= 0 \\
h_{1}&= \frac{2^{1/3}}{6} u_0 \\
h_{2}&= -\frac{2^{1/3}}{6} u_0'  \\
h_{3}&=  \frac{1}{9} u_0''\\
h_{4}&= -\frac{2^{2/3}}{36} u_0''' \\
h_{5}&= \frac{2^{1/3}}{108}\left(u_0''u_0+u_0^{(4)}-\frac{u_0^3}{3}+108u_1 \right) \\
h_{6}&=\dots
\ea
\eeq
As expected the coefficients of $h$ corresponding to power of $z$ which are not $\pm1\mbox{mod}(6)$ are
total derivatives. Moreover $h_{1}$ and $h_{5}$ are densities of the functionals (\ref{kerP0}) which are elements of the 
kernel of $P_0$.  
The Laurent series $h(z)$ plays the main role  in our  construction of the hierarchy related to $G_2^{(1)}$, we are 
indeed going to show that the knowledge of such function together with the existence of a complete set of Casimirs
of \rref{Caseq} is enough in order to compute all the solitons equations of the hierarchy.
Let us indeed first use  the invariance under the Weyl group  of $G_2$ 
of the eigenvalues of the matrix $B+\lambda A$ (which belongs to a Cartan subalgebra of $G_2$)  
(or the very expression of equation \ref{Riceqn} where only  $\lambda=z^6$ explicitly appears) 
sure that if $\psi(z)$ and $\mu(z)$ are 
respectively an eigenvector and an eigenvalue of 
the operator $-\partial_x+S+\lambda A$ then also $\psi_k(z)=\psi(e^{\frac{2\pi ik}{6}}z)$ and 
$\mu_k(z)=\mu(e^{\frac{2\pi ik}{6}}z)$ 
are for any $k=0,\dots, 5$ another pair of respectively an eigenvector and an eigenvalue.
 Moreover it is easy to show 
that for any fixed $x$ in $S^1$  the elements $\psi(e^{\frac{2\pi ik}{6}}z)$ $k=0,\dots 5$ together with the 
obviously existing ``zero''--eigenvector $\chi(z)$ form a basis of ${\Bbb C}^7$.  \par
Further using the expansion of $V^j$ in Proposition \ref{gencassol} it easy to show that any flow of the hierarchy
may be written as
\beq
\label{jhyr}
\dot{S}_j=\left[A,V_j\right].
\eeq
Then since $V^j$ is a solution of \rref{Caseq} we have   that 
$$
((V^j)_+)_x+\left[((V^j)_+,S+\lambda A\right]=-((V^j)_-)_x-\left[((V^j)_-,S+\lambda A\right]
$$
where $(\cdot)_+$ and $(\cdot)_-$ are respectively the projection on the regular and singular part of the Laurent
series $V^j$. This latter equation  implies that
$$
\left[A,V_j\right]=((V^j)_+)_x+\left[((V^j)_+,S+\lambda A\right].
$$
from which follows as usual, together with \rref{jhyr}, the bihamiltonian form of the equations of the hierarchy
\beq
\label{bihfl}
\dot{S}_j=\left[A,V_j\right]=((V^j)_+)_x+\left[(V^j)_+,S+\lambda A\right].
\eeq
It remains only to show how to rewrite these equations directly in terms of the function $h(x)$:
\begin{prop}\label{hev} The Laurent series $h(x)$ evolves as
\beq
\label{heveq}
\partial_{t_j}h=\partial_xH^{(j)},
\eeq
where the Laurent series $H^{(j)}$ called in the literature currents are 
given by $H^{(j)}=\langle v^{(0)},(V^j)_+\psi_0\rangle$.
\end{prop}
{\bf Proof.} From equations \rref{alp} with $\mu=h$ and \rref{bihfl} follows that
\beq
\label{phipsi}
(-\partial_x+S+\lambda A)\phi=h\phi-h_{t_j}\psi_0
\eeq
where 
\beq
\label{phi}
\phi=(-\partial_{t_j}+V^j)\psi_0.
\eeq
Let us now decompose $\phi$ with respect ot the basis $\chi,\psi_0,\dots,\psi_5$: $\phi=c_6\chi+\sum_{a=0}^5 c_a\psi_a$,
the equation \rref{phipsi} implies 
$$
 -c_{6x}\chi+\sum_{a=0}^5(-c_{ax}+c_ah_a)\psi_a=hc_6\chi+\sum_{a=0}^5 hc_a\psi_a-h_{t_j}\psi_1
$$
where $h_0=h$ and therefore $-c_{6x}=c_6h$ and $-c_{ax}=c_ah_a=c_ah$ for $a=1,\dots,5$, but being $c_a$ a Laurent
series in $z$ and $h$ and $h_a-h$ ($a=1,\dots, 5$) series of maximal degree $1$ these latter equation  give $c_a=0$
for $a=1,\dots, 6$. Hence $\phi=c_0\psi_0$ which taking into account \rref{phi}, the definition of $H^{(j)}$
and the normalization of $\psi_0$ implies $c_0=H^{(j)}$. Therefore $\phi=H^{(j)}\psi_0$ which plugged in \rref{phipsi}
yields  $h_{t_j}=\partial_x H^{(j)}$.
\endpf 
In the case of the Lie algebras of type $A$ the  most important property of the function $H^{(j)}$ 
is that they  can be actually computed  without using directly the Casimir $V^j$
giving a really powerful way to write down 
 all the equations of the hierarchy using simply equations  \rref{heveq}.
The matter are however much more complicated for the other affine Lie algebras, impeding a straightforward generalization
of the theory.
The difficulty arises from the fact that equation \rref{Riceqn} does not imply any more that $\lambda=z^6$ belongs 
(in the space $\mathcal{L}$ of all Laurent polynomials in $z$) 
to the linear span generated by the  Fa\`a di Bruno polynomials $H_+=\langle h^{(i)}\rangle_{i\in \Bbb N}$.
The only property which seems still survive in our setting is the observation:
\begin{prop}\label{asbeH} The Laurent series  of $H^{(j)}$ is given by
\beq
\label{asbeHeq}
H^{(j)}=(z)^j+\sum_{l\geq 1}H^j_lz^{-l}.
\eeq
\end{prop}
{\bf Proof.\ }
By the definition of $H^{(j)}$ we have 
$H^{(j)}=\langle v^{(0)},(V^j)_+\psi_0\rangle=\langle v^{(0)},V^j\psi_0\rangle-
\langle v^{(0)},(V^j)_-\psi_0\rangle=z^j-\langle v^{(0)},(V^j)_-\psi_0\rangle$, 
since $V^1\psi_0=z\psi_0$. Then since $(V^j)_-=V^j_1\lambda^{-1}+\dots$ and 
from the definition of the $h^{(k)}$ and the Riccati equation follows  that 
$\psi_0=(1,cz+O(1),(cz)^2+O(z), \dots)^T$ we have that
$$
-\langle v^{(0)},(V^j)_-\psi_0\rangle=\frac{H_1^j}{z}+\dots
$$
Hence $H^{(j)}$ has the form \rref{asbeHeq}.
\endpf
 Actually we have some fairly guesses how the theory should be modified in order to take into account at least
the   affine Lie algebras corresponding to the classical simple Lie algebras.  For instance in the case of the
affine Lie algebras $B_n^{(1)}$ the equations of the hierarchy has the form \rref{heveq} where $h$ 
 is constrained by the request that $z^{2n}h$ is in the span of the positive Fa\`a di Bruno polynomials 
and the currents $H^{(k)}$ may be directly computed as the projection on $H_+$ of $z^{n}$ 
 with respect to  the decomposition $\mathcal{L}=H_+\oplus H_-$ where  $H_-=\langle z^j\rangle_{j<0}$ together with the 
further constrains that the odd currents are linear combination of  strictly positive the Fa\`a di Bruno polynomials.
From these latters currents (in the case when $n=3$) those corresponding to the Lie algebra $G_2^{(1)}$ should be obtained
by performing a further reduction.

\end{document}